\documentclass[aps,prl,reprint,superscriptaddress,nofootinbib]{revtex4-1} 
\pdfoutput=1
\usepackage{graphicx}
\usepackage{dcolumn}
\usepackage{bm}
\usepackage[utf8]{inputenc}
\usepackage{graphics,graphicx}
\usepackage{parskip}
\usepackage{amsmath,amssymb}
\usepackage[normalem]{ulem}
\usepackage{soul}
\setstcolor{red} 
\usepackage{amsfonts}

\usepackage{multirow}
\usepackage{fancyhdr}
\usepackage{xcolor}
\usepackage{placeins}
\usepackage[title]{appendix}
\usepackage{wasysym}
\usepackage{url}
\usepackage{braket}
\usepackage{siunitx}
\usepackage{upgreek}
\usepackage{float}
\usepackage[caption=false]{subfig}

\usepackage{lipsum}
\usepackage{booktabs}
\usepackage{comment}
\usepackage{braket}
\usepackage[colorlinks=true,filecolor=.]{hyperref}
\usepackage{cleveref}
\usepackage{natbib}
\usepackage{setspace}
\usepackage{subfiles} 

\usepackage{siunitx}

\begin{document}

\title{In situ observation of non-polar to strongly polar atom-ion collision dynamics}

\author{M. Berngruber$^{\P}$} 
\affiliation{5. Physikalisches Institut, Universit\"at Stuttgart, Pfaffenwaldring 57, 70569 Stuttgart, Germany}

\author{D. J. Bosworth$^{\P}$} 
\email{dboswort@physnet.uni-hamburg.de}
\affiliation{%
  Zentrum f\"ur Optische Quantentechnologien, Universit\"at Hamburg,\\ Luruper Chaussee 149, 22761 Hamburg, Germany\\
 }%
 \affiliation{%
  Hamburg Centre for Ultrafast Imaging, Universit\"at Hamburg,\\ Luruper Chaussee 149, 22761 Hamburg, Germany\\
 }%

\author{O. A. Herrera-Sancho}
\affiliation{5. Physikalisches Institut, Universit\"at Stuttgart, Pfaffenwaldring 57, 70569 Stuttgart, Germany}
\affiliation{Escuela de F\'isica, Universidad de Costa Rica, 2060 San Pedro, San Jos\'e, Costa Rica}
\affiliation{Instituto de Investigaciones en Arte, Universidad de Costa Rica, 2060 San Pedro, San Jos\'e, Costa Rica}
\affiliation{Centro de Investigaci\'on en Ciencias At\'omicas, Nucleares y Moleculares, Universidad de Costa Rica, 2060 San Pedro, San Jos\'e, Costa Rica}

\author{V. S. V. Anasuri} 
\affiliation{5. Physikalisches Institut, Universit\"at Stuttgart, Pfaffenwaldring 57, 70569 Stuttgart, Germany}

 \author{N.~Zuber} 
\affiliation{5. Physikalisches Institut, Universit\"at Stuttgart, Pfaffenwaldring 57, 70569 Stuttgart, Germany}

\author{F. Hummel} 
\thanks{Present address: Atom Computing, Inc., Berkeley, CA, USA.}
\affiliation{%
  Max-Planck-Institute for the Physics of Complex Systems,\\ Nöthnitzer Straße 38, 01187 Dresden, Germany\\
 }%

 \author{J. Krauter} 
\affiliation{5. Physikalisches Institut, Universit\"at Stuttgart, Pfaffenwaldring 57, 70569 Stuttgart, Germany}
 
 \author{F. Meinert} 
\affiliation{5. Physikalisches Institut, Universit\"at Stuttgart, Pfaffenwaldring 57, 70569 Stuttgart, Germany}

\author{R. L\"ow} 
\affiliation{5. Physikalisches Institut, Universit\"at Stuttgart, Pfaffenwaldring 57, 70569 Stuttgart, Germany}

 \author{P. Schmelcher}
 \email{pschmelc@physnet.uni-hamburg.de}
  \affiliation{%
  Zentrum f\"ur Optische Quantentechnologien, Universit\"at Hamburg,\\ Luruper Chaussee 149, 22761 Hamburg, Germany\\
 }%
 \affiliation{%
  Hamburg Centre for Ultrafast Imaging, Universit\"at Hamburg,\\ Luruper Chaussee 149, 22761 Hamburg, Germany\\
 }%

\author{T. Pfau}
\email{t.pfau@physik.uni-stuttgart.de}
\affiliation{5. Physikalisches Institut, Universit\"at Stuttgart, Pfaffenwaldring 57, 70569 Stuttgart, Germany}
\date{\today}
\def\thefootnote{\P}\footnotetext{These authors contributed equally to this work.}

\begin{abstract}
The onset of collision dynamics between an ion and a Rydberg atom is studied in a regime characterized by a multitude of collision channels. These channels arise from coupling between a non-polar Rydberg state and numerous highly polar Stark states. The interaction potentials formed by the polar Stark states show a substantial difference in spatial gradient compared to the non-polar state leading to a separation of collisional timescales, which is observed in situ. For collision energies in the range of $k_\textrm{B}\cdot$\si{\micro K} to $k_\textrm{B}\cdot$\si{K}, the dynamics exhibit a counter-intuitive dependence on temperature, resulting in faster collision dynamics for cold -- initially ``slow" -- systems. Dipole selection rules enable us to prepare the collision pair on the non-polar potential in a highly controlled manner, which determines occupation of the collision channels. The experimental observations are supported by semi-classical simulations, which model the pair state evolution and provide evidence for tunable non-adiabatic dynamics.
\end{abstract}
\maketitle
\textit{Introduction} --- Observing, understanding and controlling individual collisions is a prerequisite for many-body physics based on atoms or molecules.
Especially in the ultracold regime, where collisions between neutral atoms can be engineered by Feshbach resonances, a high level of control is reached~\cite{chin2010feshbach}. This makes possible, for example, the study of degenerate molecular gases~\cite{de2019degenerate,duda2023transition}, Feshbach molecules~\cite{kohler2006production, thalhammer2006long} and Efimov physics~\cite{ naidon2017efimov}.
However, when it comes to collisions between charged and neutral particles reaching the same level of quantum control becomes harder since the range of interactions increases, thus requiring even lower temperatures to reach the quantum regime of scattering~\cite{weckesser2021observation, hirzler2022observation}.
More exotic collisions can be studied in systems of laser-cooled Rydberg atoms, which have the advantage of showing long-range interactions, allowing collisions to occur on larger length, slower time and lower energy scales, making them easier to observe with spatial and temporal resolution.
Even exotic bound states between a Rydberg atom and neutral ground state atoms forming ultralong-range molecules have been observed~\cite{Bendkowsky2009Observation, booth2015production, niederprum2016observation, Fey2020Ultralong-range}. 
Moreover, the complex Rydberg level structure can give rise to intriguingly rich potential energy surfaces with avoided crossings and conical intersections providing means to study effects beyond the Born-Oppenheimer approximation~\cite{Hummel2023Vibronic,Hummel2021Synthetic}.\\
More recently,  also systems combining Rydberg atoms and ions have become an active field of research~\cite{engel2018observation, schmid2018rydberg, gambetta2020long, wang2020optical, Deiss2021Long-Range, Duspayev2021Long-range, zuber2022observation}.
Here, we pursue this direction and explore the dynamical processes that lead to a multi-channel collision between an ion and a Rydberg atom. Instead of an ion trap we rely on compensating electric fields to work with free floating ions in an almost net-zero electric field environment. Our high-resolution ion microscope allows us to study collisional dynamics with both spatial- and temporal-resolution.  We are therefore not restricted to only analysing the initial and final collision partners, but may instead directly observe the dynamics as the collision unfolds.\\
\begin{figure}[t]
    \centering
    \includegraphics[width = 0.47\textwidth]{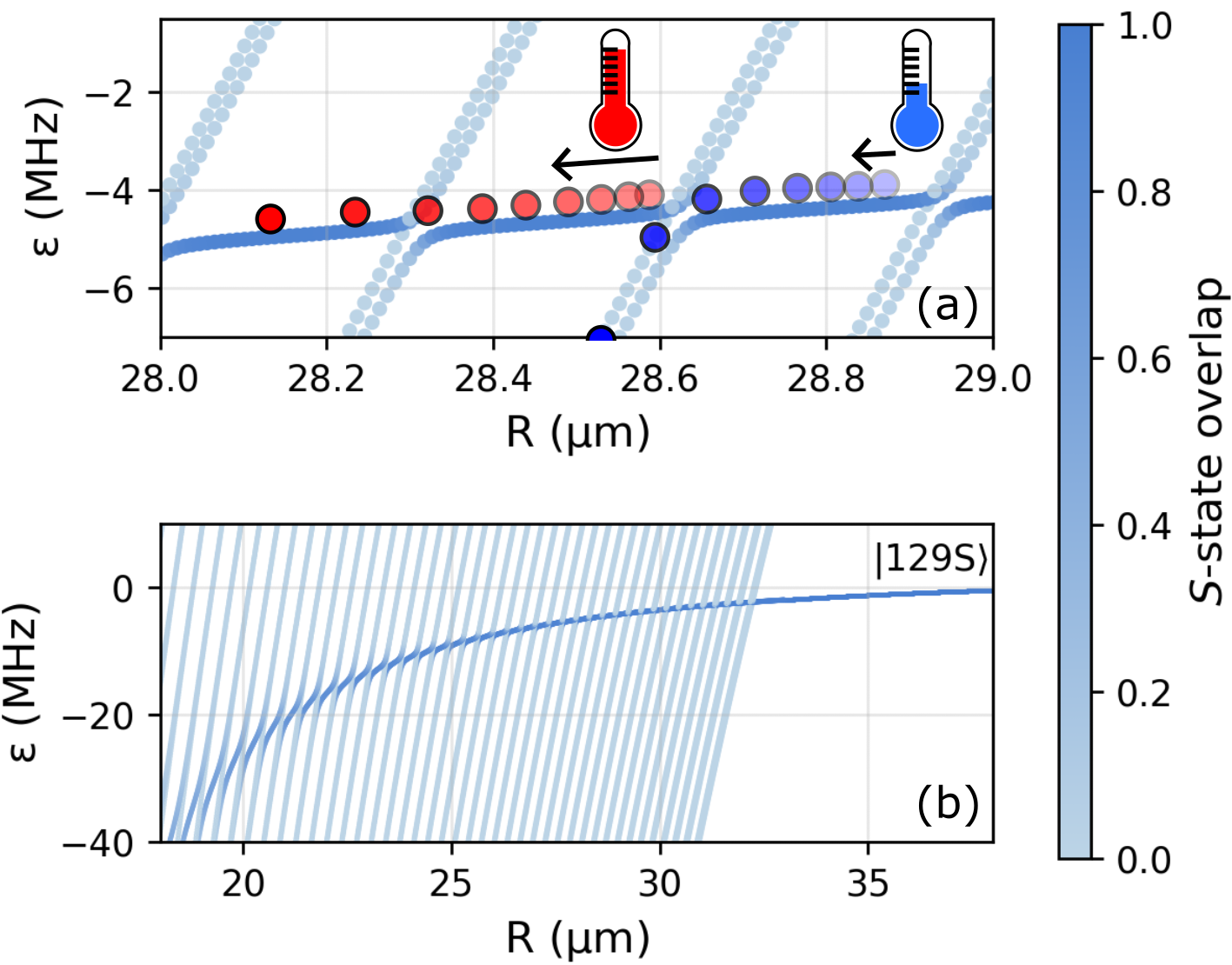}
    \caption[Model schematic.]{\textbf{Ion-Rydberg collision channels.} Adiabatic potential energy curves obtained via exact diagonalisation of the electronic Hamiltonian (see Supplementary Material). (a) Close-up around the avoided crossings near the non-polar $\ket{129S}$ atomic Rydberg state, illustrating the presence of initially slow (red) and fast (blue) collision channels. Counterintuitively, occupation of the fast channels is more probable for initially slower particles. (b) Larger-scale plot of the avoided crossings between the non-polar $S$-state and multiple strongly polar Stark states from the neighboring asymptotically degenerate hydrogenic manifold. The overlap of the electronic states with the unperturbed $\ket{129S}$ atomic Rydberg state is denoted by the colorbar. Energies are given relative to the $\ket{129S}$ atomic Rydberg state. }
    \label{fig:fig1-model}
\end{figure}
\textit{Theory} --- The polarizability of highly excited Rydberg atoms gives rise to a long-range charge-induced dipole interaction potential which is shown as a function of the internulear distance $R$ for the specific case of the $\ket{129S}$ state in Fig.~\ref{fig:fig1-model}.
These potential energy curves (PECs) are calculated by exact diagonalization of the electronic Hamiltonian $H_e = H_0 + V_I(R)$, where $H_0$ describes the unperturbed Rydberg atom and $V_I$ the ion-Rydberg interaction. This interaction term can be written in a multipole expansion, where we consider terms up to the 6$^{\rm{th}}$ order (see Supplementary Material). One can distinguish between two regions: the first region is defined in the asymptotic limit for large $R$,  where the potential of the non-polar $S$-state falls off with $1/R^4$ due to the charge-induced dipole interaction between the ion and the atomic Rydberg $S$-state; the second region s found at $R \lesssim $ \SI{33}{\micro m} for the case of $\ket{129S}$, which is nevertheless approximately 26 times larger than the size of the Rydberg orbit. Here, the ion-induced Stark shift becomes large enough that strongly polar, large angular momentum states from the neighboring $n=126$ hydrogenic manifold start to cross into the polarization potential and form a series of avoided crossings.
For Rydberg $S$-states in $^{87}$Rb, the polarization potential strictly decreases in energy while approaching the ion such  that, the two collision partners will always be accelerated toward each other.
In the direct vicinity of these avoided crossings, the Born-Oppenheimer (BO) approximation is no longer suitable to describe the dynamics properly. Instead, a non-adiabatic, semi-classical model using the Landau-Zener (LZ) formula is employed~\cite{Zener1997Nonadiabatic}. This allows to estimate the probability for an adiabatic transition to a strongly polar state at each crossing and thus can be used to predict the occupation of the different collision channels. The probability $P_{ij}$ to transition non-adiabatically from PEC $i$ to an adjacent curve $j$ is given by $P_{ij} = \exp \big(-2\pi a^2_{ij}/(\dot{R}\,\alpha_{ij})\big)$~\cite{rubbmark1981dynamical}, where $a_{ij}$ is half the energy gap at the avoided crossing and $\alpha_{ij}$ is the differential gradient between the PECs.\\
From this formula, it is clear that the probability of undergoing adiabatic dynamics at a given crossing can be experimentally tuned through the relative velocity $\dot{R}$, which is determined in an experimental setting by the atom temperature and the additional kinetic energy acquired upon falling inward on the polarization potential of the non-polar $S$-state. Therefore, systems with small relative velocities have an increased probability to follow the PEC adiabatically. In contrast, systems with high relative velocities have a larger probability to traverse the crossing non-adiabatically and thereby remain on the comparatively flat polarization potential.  Hence, each avoided crossing provides two collisional channels: one that is mostly populated by systems with low kinetic energy (\textit{cold channel}) and one that is mostly occupied by high kinetic energy systems (\textit{hot channel}), see Fig.~\ref{fig:fig1-model}. Interestingly, this leads to a counter-intuitive behaviour for the overall dynamics: if a cold, low kinetic energy system follows the PEC adiabatically, it will ultimately reach the steep strongly polar potential and thus rapidly accelerate. This results in a faster collision compared to a system with high kinetic energy, which travels along the flat $S$-state potential and experiences weaker acceleration.\\
\begin{figure}[t]
    \centering
    \includegraphics[scale = 0.89]{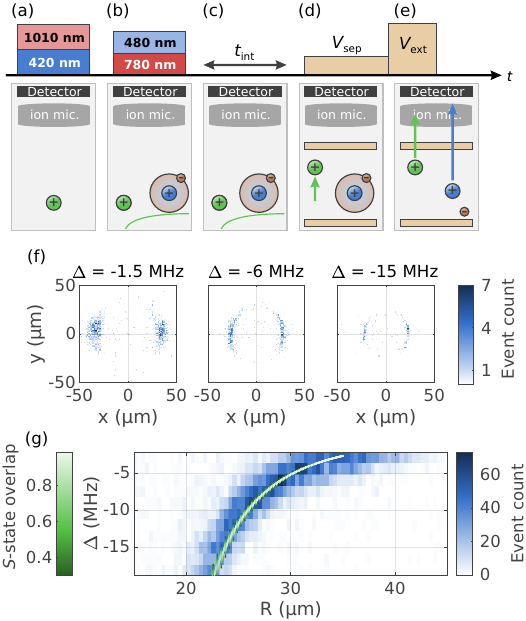}
    \caption[Experimental schematic.]{\textbf{Experimental sequence and initial state preparation.}.  (a)-(e) Schematics of the experimental sequence, which consists of the following steps: (a) two-photon ionization to create an ion, (b) two-photon Rydberg excitation in the electric field of the ion, (c) variable interaction time, (d) applying a small separation field which drags the ion along the optical axis of the ion microscope, (e) ionization of the Rydberg atom and imaging of both particles by using a large electric field. Panel (f) shows the Rydberg atom position relative to the ion (located at the origin) for different Rydberg laser detunings $\Delta$ from the bare $\ket{129S}$ state in absence of an ion. (g) Histograms of the azimuthally averaged ion-Rydberg atom distance $R$ for various detunings $\Delta$ (blue). The theoretically calculated polarization potential is shown in white to green color code, which indicates the overlap with the bare $\ket{129S}$ state (only overlaps $\geqslant 0.3$ are shown).}
    \label{fig:fig-experiment}
\end{figure}
\textit{Overview of experimental sequence} --- The charged ion-Rydberg atom system is realized in a laser-cooled rubidium cloud held in an optical dipole trap of a temperature of about \SI{20}{\micro K}. In order to minimize stray electric fields in the system, six electrodes are used to compensate stray electric fields well below \SI{100}{\micro V/cm}~\cite{veit2021pulsed}. In that way, the ions can be kept in position for the time of the experiment and no ion trap is needed. An experimental block starts with a \SI{1}{\micro s} long ionization pulse which incorporates a two-photon ionization process, providing just enough energy to overcome the ionization threshold (see Fig.~\ref{fig:fig-experiment}(a)). Next, a Rydberg atom is excited in the electric field of the ion by using a \SI{1}{\micro s} long Rydberg excitation pulse, again involving two laser beams (see Fig.~\ref{fig:fig-experiment}(b)). The detuning $\Delta$ of the upper \SI{480}{nm} excitation laser from the bare atomic state in zero field determines the initial radius $R_0$ at which the Rydberg atom is facilitated around the ion. The corresponding effective linewidth, $\dot{R}_0$ is mostly determined by the temperature of the rubidium cloud. This blue laser illuminates the atomic cloud as a thin light sheet in the horizontal direction, thus confining the system for highly excited Rydberg states to a quasi-2D plane, leading to the facilitation of Rydberg atoms located on a ring around the ion.  Afterwards a variable time $t_\mathrm{int}$ can be applied allowing the system to evolve freely.  In order to detect the two particles in a distinguishable way, we drag the ion along the optical axis of the ion microscope without displacing it in the imaging plane. To do so, two field electrodes are used to apply a weak electric field of about \SI{1.1}{V/cm}, which is small enough to not ionize the Rydberg atom (see Fig.~\ref{fig:fig-experiment}(d)). In the final detection step (Fig.~\ref{fig:fig-experiment}(e)), a large electric field of \SI{340}{V/cm} is applied to field-ionize the Rydberg atom and to accelerate both particles into the ion microscope. Due to the previous separation between the ion and the Rydberg atom, they will arrive at different times at the detector and are therefore easily distinguishable~\cite{zuber2022observation, Zou2023Observation}.\\
\textit{Results \& discussion} --- If the interaction time $t_{\rm{int}}$ in Fig.~\ref{fig:fig-experiment}(c) is set to zero, Rydberg atoms initially excited on the flat polarization potential can be directly detected at their original positions.  Excitation to the high angular momentum curves is ruled out due to negligible $D$-state overlap.  By scanning the detuning $\Delta$ of the Rydberg excitation we can spectroscopically map out the resonance condition for facilitated excitation on the interaction potential. Fig.~\ref{fig:fig-experiment}(f) shows examples of averaged in situ images of the Rydberg atom position relative to the ion, meaning that the ion is always located at the origin. Owing to the excitation in a quasi-2D plane, a symmetric ring can be observed with the ion microscope. The upper and lower part are not populated due to the finite, elongated shape of the atomic cloud. As it can be clearly seen, the distance between the ion and the Rydberg atom decreases for larger detunings as expected from the facilitation process. 
Fig.~\ref{fig:fig-experiment}(g) summarizes the result of such in situ images by integrating over the azimuthal angle and showing the data as a function of the internuclear distance $R$, which represents a direct measurement of the $\ket{129S}$ ion-Rydberg pair state potential. The blue histogram shows the experimentally obtained data, which is in good agreement with the calculated PEC displayed in green.\\
\begin{figure}[t]
    \centering
    \includegraphics[width = 0.47\textwidth]{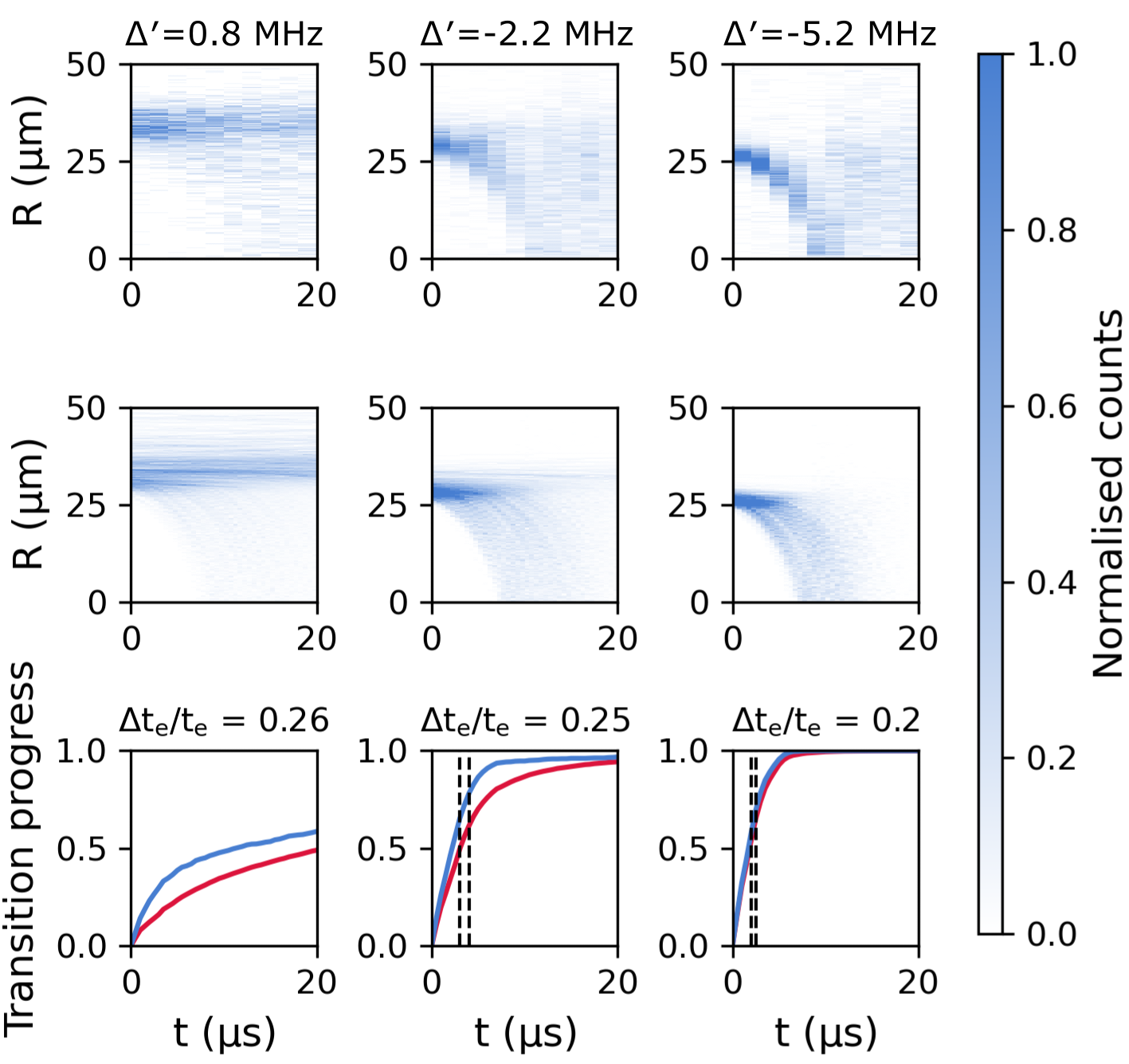}
    \caption[129S comparison.]{\textbf{Ion-Rydberg pair dynamics.} Comparison of the observed relative atomic dynamics (top row) to semi-classical LZ simulations (middle row) for different detunings of the Rydberg laser $\Delta^{\prime}$ relative to the outermost avoided crossing at $R\approx \SI{32.2}{\micro m}$. The colorbar shows the number of counts as a fraction of the total counts at $t=0$. 
    The bottom row shows the population of the fast collision channels (strongly polar states) over time for both LZ (red) and adiabatic (blue) simulations. Dashed vertical lines (not visible for $\Delta^{\prime} = 0.8$~MHz in the plotted time interval) indicate the time $t_e$ at which the population has reached $1-1/e$ (approx. 63\%). $\Delta t_e/t_e$ is the relative difference in this time between the LZ and adiabatic simulations.  For details of the simulations, see Supplementary Material.}
    \label{fig:fig-3-ion-ryd-dyn}
\end{figure}
In the next step we introduce a variable interaction time $t_{\rm{int}} > 0$ between the Rydberg excitation and the detection, during which the dynamics take place. This allows the ion-Rydberg pair to move on the interaction potential, such that the system encounters the series of avoided crossings shown in Fig.~\ref{fig:fig1-model}(b). The top row of Fig.~\ref{fig:fig-3-ion-ryd-dyn} shows results for the observed dynamics of the $\ket{129S}$ state at three different detunings. By $\Delta'$ we denote the relative detuning to the outermost avoided crossing, such that for $\Delta' > 0$ the system is initialized outside the fan of the hydrogenic manifold. Each panel represents an average over at least 6500 ion-Rydberg events on the detector. The middle row shows the result of semi-classical simulations taking into account the finite temperature, effective laser linewidth, experimental timings as well as the geometry of the setup. We model the observed ion-Rydberg pair dynamics by solving the pair's classical equation of motion along all possible collision channels. Each resulting trajectory is assigned a weight corresponding to the probability of following that particular channel, provided by the LZ formula (for further details, see Supplementary Material). We observe good overall agreement between the experimental results and the simulations. For negative $\Delta'$, one clearly observes faster dynamics in total due to the transition to steep, strongly polar states.\\
With the help of our semi-classical model, we can better understand the significance of non-adiabatic transitions in the dynamics by studying the change in population of the slow and fast collision channels over time. The bottom row of Fig.~\ref{fig:fig-3-ion-ryd-dyn} shows the population of the strongly polar states over time for our LZ simulations (red) as well as for a fully adiabatic simulation based on individual, non-coupled Born Oppenheimer PECs (blue). For all three relative detunings, the transition to the strongly polar curve is slower in the LZ simulation due to non-adiabatic transitions. To better quantify this transition we compare the time $t_e$ (dashed vertical lines) at which the population of the non-polar $S$-state curve has decayed to $1/e$ in both cases. The relative difference $\Delta t_e$ in $t_e$ between the blue and red curves consistently decreases as the Rydberg atoms are excited further inside the fan, which indicates that the pair's relative motion becomes increasingly adiabatic. This is due to the growing gap size $a_{ij}$ at the avoided crossings at smaller $R$ (see Supplementary Material). In this way the timescale of the dynamics can also be controlled via the laser detuning.\\
\begin{figure}[t]
    \centering
    \includegraphics[width = 0.47\textwidth]{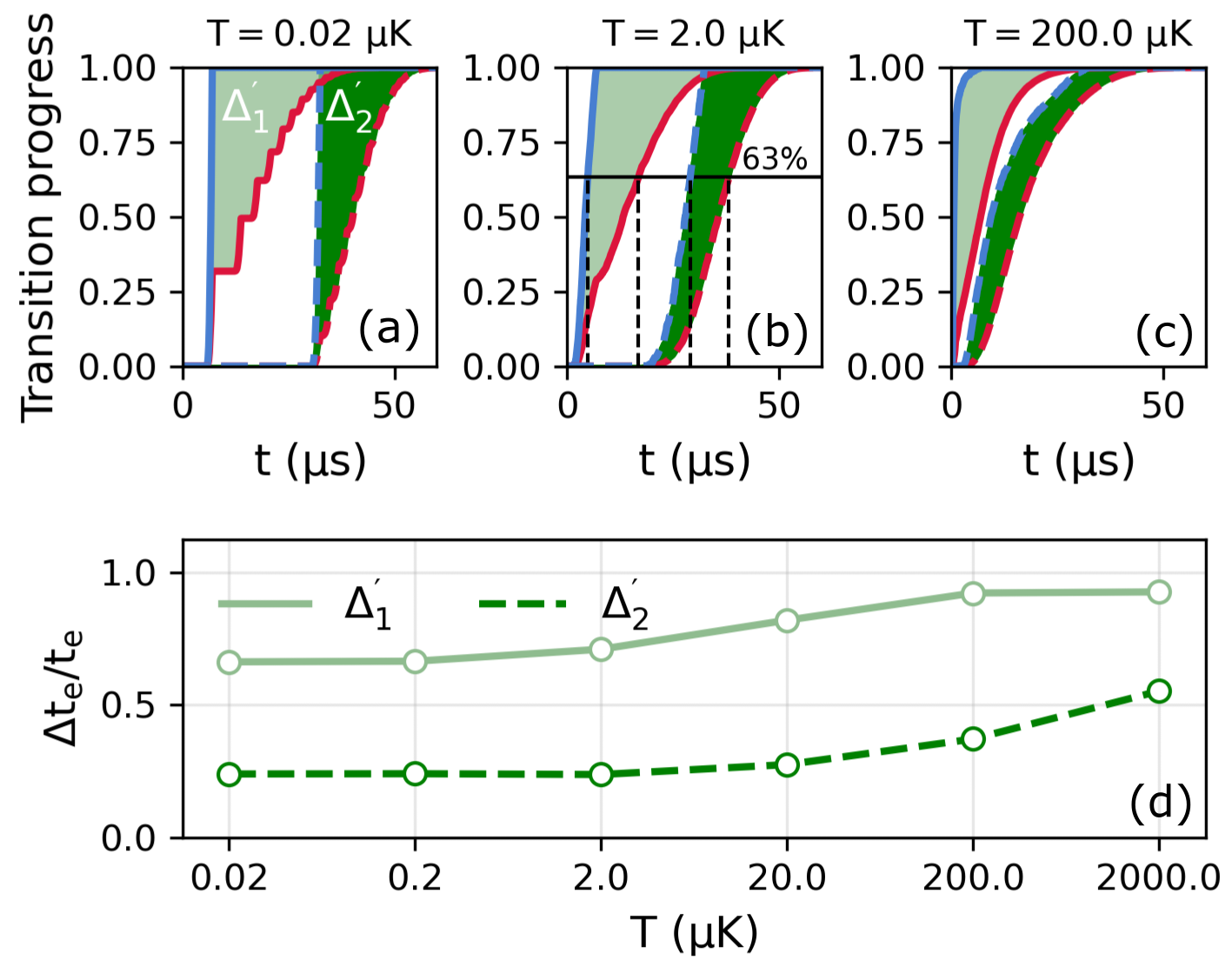}
    \caption[BBO effects.]{\textbf{Effect of the temperature on the dynamics}. Panels (a)-(c) compare the population of the fast collision channels (strongly polar states) over time for both LZ (red) and adiabatic (blue) simulations at different gas temperatures $T$. Solid (dashed) lines denote an initial laser detuning of $\Delta^{\prime}_1= 0.06$~MHz ($\Delta^{\prime}_2=0.31$~MHz) relative to the outermost avoided crossing at $R\approx$~\SI{32.2}{\micro m} (see Fig.~\ref{fig:fig1-model}(b)). Panel (d) shows the relative difference in $t_e$ between LZ and adiabatic simulations over a range of gas temperatures. $t_e$ is the time at which the polar state population reaches $1-1/e$ (approx. 63\%), indicated for the case of $T=$~\SI{2.0}{\micro K} by the dashed vertical lines in (b).}
    \label{fig:fig-4-ideal-case}
\end{figure}
Non-adiabatic transitions are further influenced by the gas temperature $T$. For high $T$, ion-Rydberg pairs have a larger probability to follow the slow collision channel due to their initially greater relative velocity. We illustrate this point in Fig.~\ref{fig:fig-4-ideal-case}, which shows the transition to fast collision channels over time at different temperatures and two different laser detunings relative to the outermost avoided crossing. The relative difference in $t_e$ increases with $T$ for both detunings (see Fig.~\ref{fig:fig-4-ideal-case}(d)), indicating that the pairs created in hotter gases spend longer on the slow collision channel created by the non-polar $S$-state. Pairs gain kinetic energy as they fall inward along the polarization potential. For $\Delta^{\prime}_1 = 0.06$~MHz, this gain in kinetic energy corresponds to approximately \SI{0.4}{\micro K}$\cdot k_{\textrm{B}}$ of thermal energy, whilst $\Delta^{\prime}_2 = 0.31$~MHz is equivalent to \SI{2.0}{\micro K}$\cdot k_{\textrm{B}}$. This additional heating accounts for the negligible change in $\Delta t_e/t_e$ for temperatures below \SI{2}{\micro K} (Fig.~\ref{fig:fig-4-ideal-case}(d)).\\
Non-adiabatic transitions will also play a more significant role in the dynamics at higher $n$ as a result of the narrowing avoided crossings, whose gap sizes $a_i$ follow a power-law decay with $n$ (see Supplementary Material). The signal of non-adiabatic transitions can thus be enhanced by probing dynamics at larger $n$.\\
\textit{Summary \& outlook} --- In conclusion, we studied the onset of collisional dynamics between an ion-Rydberg pair in a regime of multiple coupled channels with different characteristic collision timescales. We were able to describe the experimentally observed dynamics with the help of a LZ model. Further, the simulations show that the collisional dynamics can not only be tuned by the initial distance but should also be tunable by other parameters like the principal quantum number.\\
Our work has explored the role of non-adiabatic effects in ion-Rydberg collisions and lays the foundation for future explorations of beyond Born-Oppenheimer physics with Rydberg atoms, such as molecular dynamics in the presence of conical intersections. Precisely understanding non-adiabatic couplings in complicated potential energy landscapes is also a key ingredient to better predict the lifetime of Rydberg molecules such as macrodimers or ion-Rydberg molecules~\cite{Hummel2021Synthetic}. In this current experimental realization, the ion-Rydberg complex was photoassociated directly out of a trapped gas of $^{87}$Rb. Future experiments might consider using individually trapped atoms in a tweezer setup, which would offer more precise control over the initial separation of the ion-Rydberg pair, thereby further improving the starting conditions of the collision.
\section*{Acknowledgements}
This work received funding from 
the DFG as part of the SPP 1929 "Giant Interactions in Rydberg Systems (GiRyd)" [Project No. Pf 381/17-1 and No. Pf 381/17-2], 
and got further funding from
the Cluster of Excellence “Advanced Imaging of Matter” of the Deutsche Forschungsgemeinschaft (DFG)-EXC 2056, Project ID No. 390715994. 
Furthermore we are supported by 
the European Research Council (ERC) under the European Union’s Horizon 2020 research and innovation programme (Grant Agreement No. 101019739-LongRangeFermi). 
F.M. received funding from the the Federal Ministry of Education and Research (BMBF) under the grant CiRQus. 
O.A.H.S. acknowledges great support from the Alexander von Humboldt Foundation. 
%
\appendix
\section{Supplementary Material}
\subsection{A --- Experimental sequence}\label{ssec:appendix-exp-sequence}
The experimental sequence starts with the preparation of a cold $^{87}$Rb cloud. The atoms originate from an effusive oven source and are subsequentially cooled by employing a Zeeman slower and trapped in a magneto-optical trap (MOT) in a separate MOT-chamber. After a \SI{15}{ms} long compressed MOT phase followed by a \SI{25}{ms} optical molasses phase, the atoms are transferred to a movable dipole trap which transports the atoms into the science chamber below the ion microscope. At this point the atoms have a temperature of about \SI{20}{\micro K}. In such a sample several thousand experiments can be performed before a new atomic sample has to be loaded.\\
A single experimental block starts with the creation of an ion by using a two-photon ionization process. The lower transition is realized by a \SI{420}{nm} laser beam which is detuned by \SI{80}{MHz} from the intermediate $\ket{6P_{3/2}, F = 3}$ state. The beam is shone into the atomic cloud in the horizontal direction and has a 1/$e^2$ waist of $w_{420} = \SI{7}{\micro m}$. The upper laser is operated at a wavelength of \SI{1010}{nm}, providing just enough energy to overcome the ionization threshold. By shining this laser vertically in the experimental chamber with a small 1/$e^2$ waist of $w_{1010} = \SI{3.2}{\micro m}$ one can create a reasonably good ion spot. Both lasers are simultaneously on for \SI{1}{\micro s}. In the subsequent step the Rydberg atom is excited in the electric field of the ion, which also takes place within \SI{1}{\micro s}. Here, a two-photon process is employed once-again, involving a \SI{780}{nm} laser, \SI{250}{MHz} detuned from the intermediate $\ket{5P_{3/2},F=3}$ state. The second laser is operated at around \SI{480}{nm}, depending on the desired target Rydberg state. While the \SI{780}{nm} laser is much larger than the atomic cloud and shone in from below, the \SI{480}{nm} beam forms a thin light sheet and is introduced horizontally into the chamber. Thus, the system can be approximated by quasi-2D plane for large Rydberg states.\\
In the following step a variable interaction time $t_\mathrm{int}$ can be introduced before the two particles are detected. To distinguish them, a small electric field is applied along the $z$-axis, which drags the ion along the optical axis of the ion microscope but does not yet ionize the Rydberg atom. When the Rydberg atom is field-ionized in the next step by applying the large extraction field of \SI{340}{V/cm}, the ion and the Rydberg atom have different starting positions and therefore will arrive at different times on the detector.
\begin{figure}[t]
    \centering
    \includegraphics[width = 0.475\textwidth]{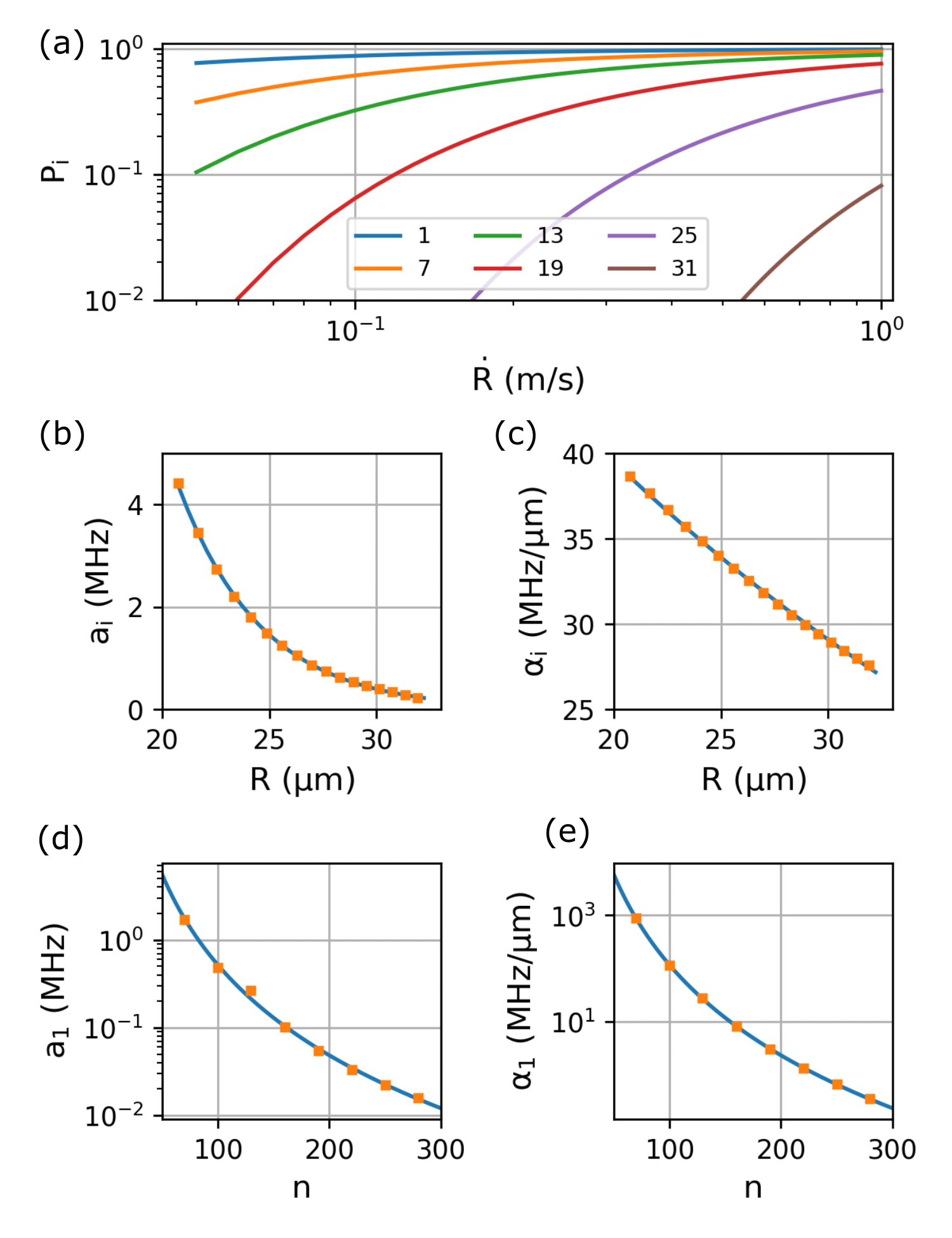}
    \caption[Supplementary Fig.]{\textbf{Scaling of Landau-Zener transition parameters.} (a) Transition probability as a function of the relative internuclear speed $\dot{R}$ for a selection of avoided crossings in the ion-129$S$ Rydberg system, where the crossings are indexed such that 1 refers to the outermost avoided crossing.  (b),(c) Scaling of the gap size $a_i$ and differential gradient $\alpha_i$ for each crossing in the ion-129$S$ Rydberg system. Solid lines represent fits to exponentially decaying functions of the form $y = A \exp(-b x)$. We find fit values of $b = 0.258\pm0.002$ and $b = 0.031\pm0.001$ for $a_i$ and $\alpha_i$, respectively. (d),(e) Scaling of the gap size $a_1$ and differential gradient $\alpha_1$ of the outermost avoided crossing with varying principal quantum number $n$. Solid lines represent fits to power-law-decaying functions of the form $y = A x^{-b}$. We find fit values of $b = 3.422\pm0.075$ and $b = 5.641\pm0.0.009$ for $a_1$ and $\alpha_1$, respectively.  In (b)-(e), the fits have a root-mean-square error of less than 10\%.}
    \label{fig:LZ-scaling}
\end{figure}
\subsection{B --- Details on the Landau-Zener model for ion-Rydberg pair dynamics}\label{ssec:appendix-theory}
We employ a semi-classical model for the ion-Rydberg pair dynamics, in which the pair's classical motion is determined along the $N$ relevant open collision channels. In the case of $\ket{129S}$, $N=34$. The open channels are obtained from the adiabatic potential energy curves (PECs) of the system's electronic Hamiltonian $H_e$, which reads: 
\begin{equation}~\label{eq:e-ham}
  H_e = H_0 + V_I(R).
\end{equation}
Here, $H_0$ denotes the unperturbed Rydberg atom and $V_I$ describes the interaction between the ion and the Rydberg atom. We are interested in ultralong-range internuclear separations $R$ at which the overlap between the species' charge distributions vanishes, enabling $V_I$ to be approximated using a multipole expansion~\cite{Duspayev2021Long-range,Deiss2021Long-Range}:
\begin{equation}\label{eq:ion-ryd-int}
        V_{I}(R) = -\sum_{\lambda=1}^{\infty} \sqrt{\frac{4\pi}{2\lambda + 1}}\,\frac{r^\lambda}{R^{\lambda+1}} Y_{\lambda}^{0}(\theta,\phi),
\end{equation}
where $r$ is the position of the Rydberg electron with respect to the core, $Y_{\lambda}^{0}(\theta,\phi)$ are spherical harmonics which are functions of the Rydberg electron's angular position $(\theta,\phi)$ and $\lambda$ denotes the order of the multipole expansion. The above expansion in~\eqref{eq:ion-ryd-int} is valid in the regime $R\gg r$. The adiabatic PEC $\{\varepsilon_i(R)\}$ are determined by evaluating eigenvalues of the electronic Hamiltonian in Eq.~\eqref{eq:e-ham} over a range of internuclear separations $R$ using the \textit{pairinteraction} program \cite{Weber2017}, which truncates the series in Eq.~\eqref{eq:ion-ryd-int} at $\lambda = 6$.\\
We are particularly interested in adiabatic PEC which correspond asymptotically to weakly-polar atomic Rydberg states. At large internuclear separations, the non-polar $\ket{nS}$ state acquires an induced dipole moment and the leading-order correction to its energy is given by $\varepsilon_S\propto -n^7/R^4$. In the Rydberg series of $^{87}$Rb,  the quantum-defect-split $\ket{nS}$ state lies below the $n-3$ degenerate hydrogenic manifold. The states in this manifold have an angular momentum of $l > 3$ and acquire a permanent dipole moment at large $R$ due to $l$-mixing, which makes them strongly polar and gives them a leading-order energy correction of $\varepsilon_l\propto \pm n^2/R^2$.\\
The competing $R$-scaling of $\varepsilon_S$ and $\varepsilon_L$ means that the attractive branches of the Stark-split high-$l$ states eventually become near-degenerate with the $\ket{nS}$ state at sufficiently low $R$, giving rise to a series of avoided crossings (see Fig.~\ref{fig:fig1-model}(b) in the main text). We refer to this region of internuclear separations $R_c$ as the Inglis-Teller limit, which was defined originally in studies on plasmas as the point at which Stark broadening is sufficiently strong to mix states of different $n$~\cite[p.~75]{Gallagher1994Rydberg}. Based on the asymptotic behaviour of the non-polar and polar PEC in our system, the onset of this regime should scale roughly as $R_c \propto n^{5/2}$, which is equivalent to a critical electric field strength of $E \propto n^{-5}$  and agrees with earlier results~\cite{Inglis1939Ionic}.\\
In the vicinity of these avoided crossings, the adiabatic Born-Oppenheimer approximation is no longer strictly valid and in general it becomes necessary to employ a coupled-channel formalism for modelling the vibrational dynamics~\cite{Koeppel1984Multimode},  often at greater computational cost. In contrast, the Landau-Zener (LZ) formula~\cite{Zener1997Nonadiabatic} provides a straightforward semi-classical approach for modelling the dynamics of vibrational degrees of freedom in the vicinity of avoided crossings and has been previously applied within the context of ultralong-range Rydberg molecules for predicting electronic transitions due to non-adiabatic couplings~\cite{Schlagmueller2016Ultracold}. The LZ formula states that for a wavepacket moving with speed $\dot{R}$ toward an avoided crossing between two coupled PECs $\varepsilon_i(R)$ and $\varepsilon_j(R)$, the system may undergo a non-adiabatic transition between the channels with probability 
\begin{equation}\label{eq:landau-zener}
   P_{ij} = \exp \bigg(-2\pi\frac{a^2_{ij}}{\dot{R}\,\alpha_{ij}}\bigg).
\end{equation}
Here $a_{ij}$ is half the energy gap at the avoided crossing and $\alpha_{ij}$ is the differential gradient between the curves. For brevity, we re-write Eq.~\eqref{eq:landau-zener} in terms of a single index $i$ for each avoided crossing $P_{ij}\rightarrow P_i$, where $i=1$ corresponds to the outermost (i.e. large $R$) avoided crossing in the ion-Rydberg PEC.\\
As shown in Fig.~\ref{fig:LZ-scaling}(a) for Rydberg atoms excited to the $\ket{129S}$ state, the ion-Rydberg pair has a finite probability to transition between PECs at the avoided crossings, which opens up a multitude of potential collision channels. The transition probability rises with increasing relative velocity $\dot{R}$ and decreases drastically for avoided crossings at smaller internuclear separations. This is primarily due to the increasing gap size $a_i$ at small $R$, which can be seen in Fig.~\ref{fig:LZ-scaling}(b). The quantities featured in Eq.~\eqref{eq:landau-zener} show significant variation with principal quantum number $n$. The variation in the gap size and gradient at the outermost crossing ($i=1$) with $n$ is shown in Fig.~\ref{fig:LZ-scaling}(d) and \ref{fig:LZ-scaling}(e). We expect that the quadratic depedence of $a_i$ in Eq.~\eqref{eq:landau-zener} will ensure that its variation dominates the change in transition probabilities compared to variations in the differential gradient $\alpha_i$. Thus, the probability for non-adiabatic transitions will increase with $n$. \\
We model the dynamics of the ion-Rydberg pair starting on the $\ket{129S}$ PEC and account for non-adiabatic transitions between collision channels via the LZ formula. The initial conditions $R(0) = R_0$ and $\dot{R}(0) = \dot{R}_0$ of the pair's equation of motion are determined from the Rydberg laser detuning $\Delta$ and the gas temperature $T$, respectively.
To ensure a fair comparison with the experimental measurements, it is necessary to account for the finite effective linewidth of the Rydberg excitation laser $\delta$, which broadens the initial separation distribution of the ion-Rydberg pairs $\{R_0\}$. In addition, various time delays are present during the preparation and extraction steps of the experimental sequence due to the finite operation time of the excitation lasers and the time of flight of the ions to the multi-channel plate detector, which have also been accounted for in our simulations.
\begin{figure}[t]
    \centering
    \includegraphics[width = 0.475\textwidth]{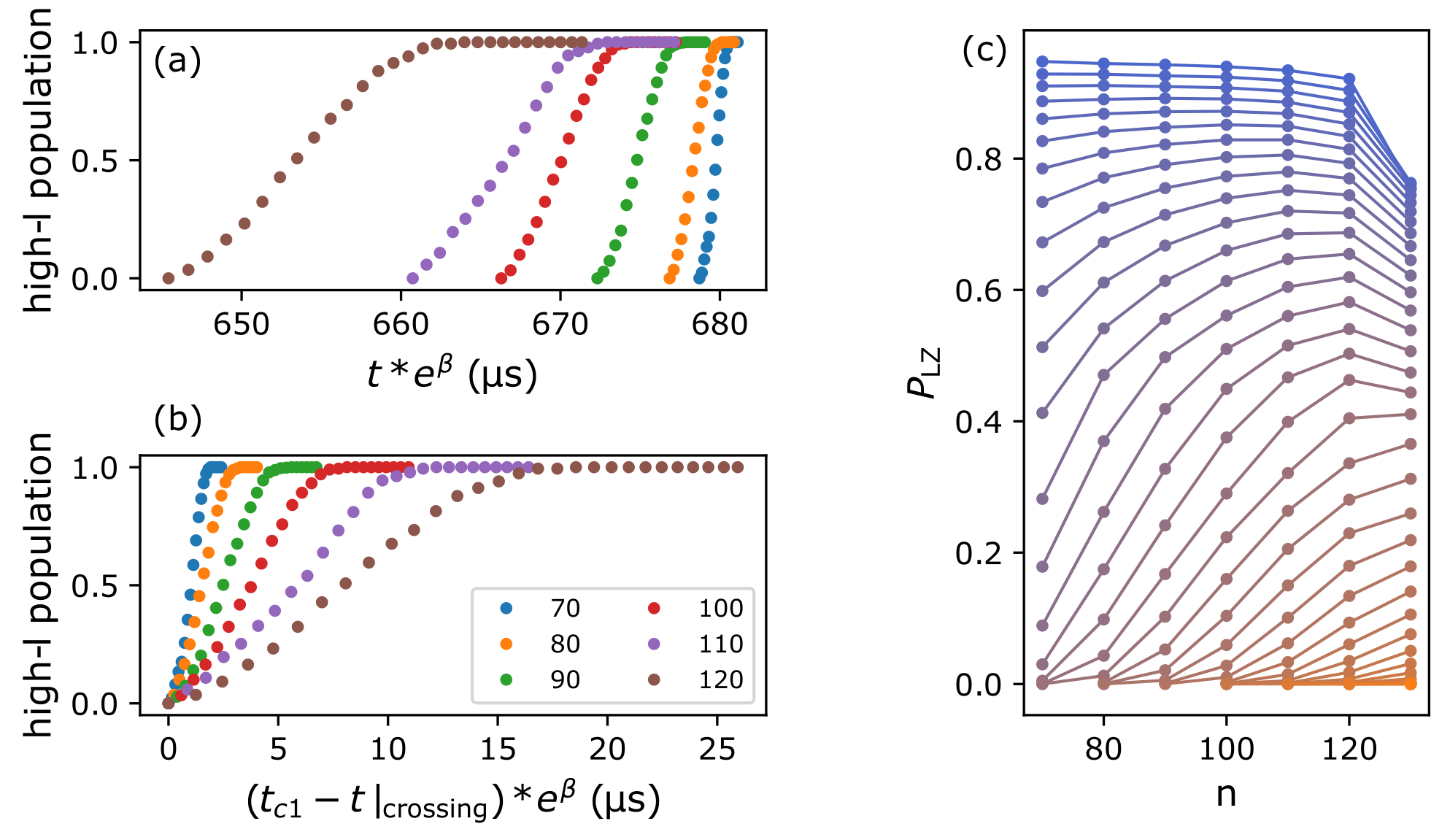}
    \caption[129S comparison.]{\textbf{Dynamics for different principal quantum numbers $\bm{n}$.} (a) The population of the  strongly polar high-$l$ states is plotted on a rescaled time axis, where $\beta(n)$ is chosen in such a way that the dynamics on the bare $S$-state potential collapse on the $n=70$ curve for all $n$. (b) For better comparison the arrival time of the first crossing is set to zero. (c) Transition probability as a function of $n$. Colors represent different avoided crossings - ranging from the outermost crossing (blue) to the innermost crossing (orange). Probabilities corresponding to the same crossing are connected by lines to guide the eye.}
    \label{fig:nScaling}
\end{figure}
\subsection{C --- \textit{n}-scaling of non-adiabatic transition probabilities}\label{ssec:appendix-n-scaling}
We now investigate the tunability of the LZ transition probabilities with principal quantum number $n$. For the initial separation we choose a fixed distance of \SI{33}{\micro m}, such that for all $n$ the Rydberg atom is excited onto the $\ket{nS}$ PEC before it crosses into the fan of Stark-split strongly polar states.  For simplicity, we set the initial velocity and the effective laser linewidth to zero and consider a one-dimensional case. \\
The duration of the dynamics varies considerably with $n$ due to the changing length scale and gradient of the PEC (see Fig.~\ref{fig:LZ-scaling}(d) and \ref{fig:LZ-scaling}(e)).  Consequently,  to be able to easily compare the relative adiabaticity of the different dynamics,  we rescale the time-axis by a factor $e^\beta$, where $\beta$ is chosen for each $n$ in such a way that the dynamics on the bare $S$-state potential collapse onto the same trajectory.  Here, we choose $n = 70$ as the reference trajectory. Therefore, $\beta(n{=}70) = 0$ and we find that $\beta$ strictly increases for larger $n$, reaching $\beta(n{=}130) \approx 2.2$ for the largest considered principal quantum number.\\
Fig.~\ref{fig:nScaling}(a) shows the fraction of systems that populate the strongly polar high-$l$ states as a function of the rescaled time, where each point corresponds to the time at which a crossing is reached. For higher $n$, the avoided crossings are reached earlier and as a result the population starts to increase at earlier times. In Fig.~\ref{fig:nScaling}(b), the arrival time at the first crossing is set to zero for all curves, revealing the difference of the adiabaticity inside the fan of the hydrogenic manifold for different $n$. This is mainly dominated by the density of the crossings since, the LZ probability at each individual crossing does not change drastically with $n$ for the first 10 crossings, as can be seen in Fig.\ref{fig:nScaling}(c)).  In this figure,, we plot the LZ probabilities for all individual crossings, starting with the outermost crossing shown in blue to the deepest considered crossing shown in orange. The deviation from the overall trend at $n=130$ is due to the fact that the atoms are initialized very close to the first crossing for this state.\\
\begin{figure}[t]
    \centering
    \includegraphics[width = 0.475\textwidth]{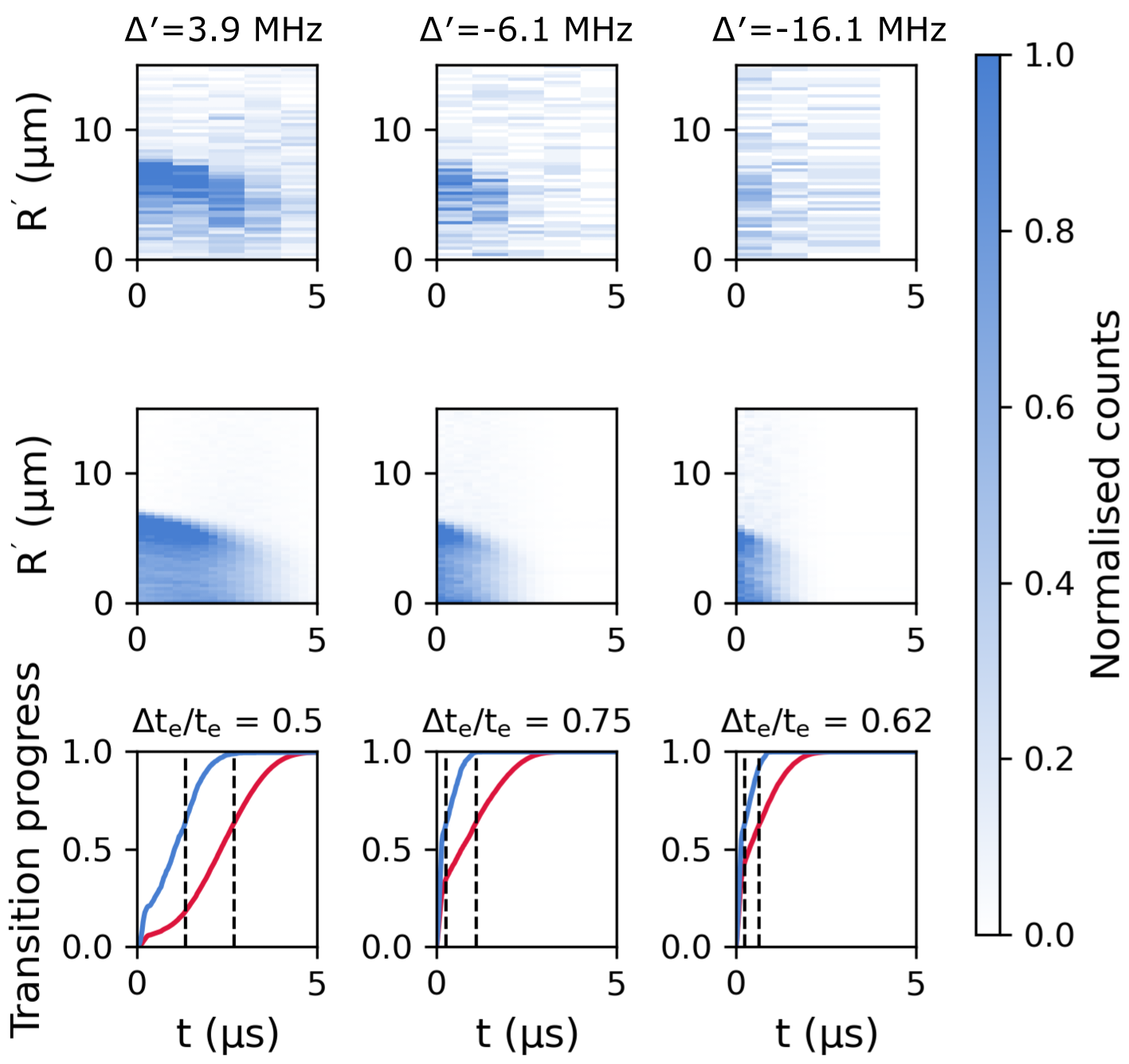}
    \caption[70S comparison.]{\textbf{Ion-Rydberg pair dynamics.} Comparison of the observed (projected) relative atomic dynamics $R^{\prime}(t)$ (top row) to semi-classical simulations (middle row) for different detunings of the Rydberg laser $\Delta^{\prime}$ relative to the $\ket{70S}$ atomic Rydberg state. The bottom row shows the population of the fast collision channels over time for both LZ (red) and adiabatic (blue) simulations. Dashed vertical lines indicate the time $t_e$ at which the population has reached $1-1/e$ (approx. 63\%). $\Delta t_e/t_e$ is the relative difference in this time between the LZ and adiabatic simulations.}
    \label{fig:70S-ion-ryd-dyn}
\end{figure}
\subsection{D --- Ion-Rydberg pair dynamics for the $\ket{70S}$ state}\label{ssec:further-dynamics}
We also observed ion-Rydberg dynamics for the $\ket{70S}$ atomic Rydberg state and the results are shown in Fig.~\ref{fig:70S-ion-ryd-dyn}.  For $\ket{70S}$, the Inglis-Teller regime manifests at smaller internuclear separations than $\ket{129S}$ (cf. $R_c\approx$~\SI{6.6}{\micro m} to $R_c\approx$~\SI{32.2}{\micro m}) due to the lower principal quantum number of the Rydberg excitation. Moreover, ion-Rydberg collision pairs at $\ket{70S}$ have fewer open channels than at $\ket{129S}$ due to the smaller number of degenerate high-$l$ states in the neighboring $n=67$ manifold.\\ 
Additionally, in contrast to the measurements taken for $\ket{129S}$, no 2D confinement was employed during the experimental sequence and consequently the dynamics unfold in 3D. As a result, the measurements show the \textit{projection} of the internuclear separation $R$ onto the micro-channel plate, denoted by $R^{\prime}$, instead of the true value of $R$. This has the consequence that the ion-Rydberg separation distribution is extremely broad -- even for time $t=0$ -- as can be seen in the top and middle rows of Fig.~\ref{fig:70S-ion-ryd-dyn}.\\
As the ion-Rydberg pairs travel inward to small internuclear separations, they will eventually reach an extreme regime in which the Stark-splitting of neighboring and next-neighboring hydrogenic manifolds becomes far greater than their field-free energetic separation. Thus, the electronic states of the Rydberg atom become highly mixed and an accurate description of the physics becomes increasingly difficult due to the sheer multitude of open channels. We do not aim at a description of the dynamics at such small internuclear separations and instead focus on capturing the physics at intermediate separations at the onset of the Inglis-Teller regime. Therefore, we introduced a short-distance cut-off to the simulations which removes particles once they have crossed a certain threshold separation,  equal to \SI{0.1}{\micro m} for the case of $\ket{70S}$.\\
When comparing the LZ and adiabatic simulations in the bottom row of Fig.~\ref{fig:70S-ion-ryd-dyn}, we see that the absolute difference in the transition times $\Delta t_e$ to the strongly polar curves decreases with decreasing detuning $\Delta^{\prime}$ -- indicating, as expected, the increased frequency of non-adiabatic transitions at higher detuning. However, unlike the simulations carried out for $\ket{129S}$, there is no clear trend in the relative difference of these transition times. We attribute this partially to the effective linewidth of the excitation laser $\delta =$ \SI{3.4}{MHz}, which leads to broadening of the initial separation distribution $\{R_0\}$ of the collision pairs. Whilst this broadening is also present at $\ket{129S}$, the effect is amplified in the case of $\ket{70S}$ due to the smaller length scales at which the dynamics occur.\\  
\end{document}